\documentclass[aps,prx,reprint,superscriptaddress]{revtex4-2}

\usepackage{graphicx}
\usepackage{amsmath,amssymb,amsfonts}
\usepackage{bm} 
\usepackage{hyperref} 
\usepackage{color}
\usepackage{xcolor}
\usepackage{soul}
\setstcolor{red}
\usepackage{booktabs}
\usepackage[utf8]{inputenc}
\usepackage{comment}
\usepackage{subcaption}

\usepackage{multirow}
\usepackage{algorithm}
\usepackage{algorithmicx}
\usepackage{algpseudocode}
\usepackage{listings}
\usepackage{appendix}
\usepackage{ragged2e}
\usepackage{physics}

\usepackage{enumitem}
\setlist{leftmargin=3.5mm}

\usepackage{placeins} 

\usepackage{xcolor}
\usepackage[normalem]{ulem}
\usepackage{tikz}
\usetikzlibrary{decorations.markings}
\usetikzlibrary{calc}
\usetikzlibrary{arrows.meta, positioning}

\begin{document}

\preprint{APS/123-QED}


 \author{Ayush Asthana}
 \email{ayush.asthana@und.edu}
 \affiliation{Department of Chemistry, University of North Dakota, Grand Forks, ND 58201, USA}
 
\title{Exact and Tunable Quantum Krylov Subspaces via Unitary Decomposition}
\begin{abstract}
Quantum Krylov subspace methods can extract ground and excited states by diagonalizing the Hamiltonian in a compact variational space. In practice, these spaces are almost always generated by real or imaginary time evolution, forcing a timestep trade-off between dynamical accuracy and basis collapse and often producing ill-conditioned overlap matrices that stall convergence. Here we introduce Quantum Krylov using Unitary Decomposition (QKUD), a time-evolution-free construction that maps Hamiltonian powers to implementable unitaries via the Hermitian transform $\sin(\epsilon H)/\epsilon$. QKUD reduces to the exact Hamiltonian-power Krylov recursion as $\epsilon\rightarrow0$, while finite $\epsilon$ provides a controllable deformation that tunes subspace geometry and improves conditioning. Across molecular active-space benchmarks and a frustrated 2D J1-J2 Heisenberg model, QKUD reproduces exact-Krylov convergence in well-conditioned regimes and systematically restores variational improvement when both exact Krylov and time-evolution Krylov stagnate. These results identify overlap conditioning, instead of time-evolution fidelity, is the key resource for robust quantum Krylov simulation and provide a resilient way forward for accurate quantum simulation of challenging quantum many-body problems.
\end{abstract}

\maketitle

\section{Introduction}
Rapid advances in quantum hardware and algorithms~\cite{alexeev2025perspective,tilly2021variational,mcardle2020quantum,cerezo2021variational,peruzzo2014variational,magann2021pulses,kandala2017hardware,cao2019quantum,fedorov2022vqe}
have led to early fault-tolerant demonstrations and first experiments on emerging platforms~\cite{google2025observation,robledo2024chemistry,barison2025quantum,danilov2025enhancing,yoshida2025auxiliary},
yet it remains unclear which algorithmic paradigms can deliver decisive advantage on scientifically relevant many-body problems under realistic resource constraints~\cite{babbush2025grand}.
Quantum phase estimation (QPE)~\cite{kitaev1995quantum,bauer2020quantum,lee2023evaluating} is a flagship fault-tolerant approach, but its resource demands and the initial-state problem~\cite{lee2023evaluating} motivate complementary strategies.
Near-term variational methods such as ADAPT-VQE~\cite{grimsley2019adaptive,ramoa2025reducing} and other directions including dissipative engineering and quantum control~\cite{lin2025dissipative,Meitei2020,asthana2022minimizing}
are promising but can incur substantial optimization or measurement overhead.
In this landscape, quantum Krylov subspace algorithms~\cite{cortes2022quantum,nandy2025quantum,kirby2023exact,tkachenko2024quantum,lee2024sampling,shen2023real}
(see Ref.~\cite{motta2024subspace} for a review) offer a meaningful middle ground by extracting eigenstates from compact variational subspaces whose basis vectors can be represented exactly on a quantum processor~\cite{rohwedder2011analysis}. They have also been recently used for error mitigation in large systems~\cite{fischer2025large}
A broad family of Krylov constructions has been proposed, including QLanczos~\cite{motta2020determining}, QRTE and related filter-diagonalization methods~\cite{parrish2019quantum,cohn2021quantum,oumarou2025molecular,klymko2022real},
QITE-based approaches~\cite{mcardle2019variational,motta2020determining,yeter2020practical}, multireference and power-method variants~\cite{stair2020multireference,zhang2024measurement,seki2021quantum,qdavidson},
and recent sampling-diagonalization implementations~\cite{yu2025quantum,yoshioka2025krylov}.
However, almost all practical quantum Krylov methods construct the basis by real or imaginary time evolution, introducing a timestep $\Delta t$ that must be chosen a priori.
This creates an intrinsic tradeoff: small $\Delta t$ improves dynamical fidelity but drives successive basis vectors toward near-linear dependence (basis collapse), while larger $\Delta t$ can alleviate collapse at the cost of system-dependent distortions of the Krylov span.
Consequently, aside from resource-intensive exact constructions such as Chebyshev-polynomial Krylov~\cite{kirby2023exact}, current time-evolution Krylov approaches are not strictly exact and can stagnate or become highly parameter sensitive with no standard way of finding a parameter that universally works. 

Here we address this bottleneck by constructing quantum Krylov subspaces without explicit time evolution.
Building on unitary decomposition techniques~\cite{schlimgen2021quantum,schlimgen2022quantum}, we map Hamiltonian powers to unitary operations on a quantum device and introduce Quantum Krylov using Unitary Decomposition (QKUD).
QKUD recovers strict Krylov recursion in the limit $\epsilon\to0$, while finite $\epsilon$ provides a controlled deformation of the subspace geometry (conditioning and linear independence) rather than discretizing physical time.
This decoupling eliminates timestep-selection ambiguity and avoids the basis-collapse pathology inherent to time-stepped constructions.
Unlike Chebyshev polynomial Krylov methods, which require high-degree unitary polynomial approximations to $H$ (hence deep circuits), QKUD achieves exactness in principle and also offers a continuous interpolation between exact and deformed subspaces.
We show across molecular benchmarks and a many-body spin model that QKUD mimics exact Krylov convergence at low $\epsilon$ and exactly solves the problem when the exact Krylov is well behaved. In cases where exact Krylov saturates early, increasing $\epsilon$ in periods of $\frac{(2n+1)\pi}{2||H||}$ or $\frac{(n)\pi}{||H||}$ provides a systematic way of deforming the subspaces to break through the ill-conditioning of the overlap matrix. 
This approach suggests a new paradigm for quantum algorithms by treating subspace geometry as a controllable resource, potentially improving robustness across many applications and associated algorithms.

\section{Theory}\label{theory}

Krylov algorithms build a variational subspace by repeatedly applying an operator derived from the Hamiltonian to an initial state. 
Writing $|\Psi_0\rangle=\sum_j c_j|E_j\rangle$, one has $\hat H^n|\Psi_0\rangle=\sum_j c_j E_j^n |E_j\rangle$, so successive Krylov vectors contain increasingly high powers of the eigenvalues. 
With a suitable shift/rescaling (so the ground-state eigenvalue is extremal in magnitude), these powers act as a spectral filter that effectively magnifies the ground-state contribution relative to excited-state components, enabling rapid convergence upon subspace diagonalization.

\subsection{QKUD: Quantum Krylov using Unitary Decomposition}
In this section, we outline the theoretical formulation of QKUD method. The key idea is to construct the Krylov basis using a Hermitian combination of unitary operators derived from the Hamiltonian, instead of relying on approximate time-evolution. For a Hermitian Hamiltonian $\hat{H}$ (such as a molecular Hamiltonian), we define the unitary operator
\begin{align}
    X = ie^{-i\epsilon\hat{H}},
\end{align}
where $\epsilon$ is a small real deformation parameter. Note that $X$ is unitary (up to the global phase $i$), so $X+X^\dagger$ is a Hermitian operator. Starting from an initial state $|\Psi_0\rangle$, we generate the Krylov subspace by repeatedly applying $X+X^\dagger$ (developed using unitary decomposition technique in Ref. ~\cite{schlimgen2021quantum}). The first Krylov basis state is
\begin{align}
    |\Psi_1\rangle = \frac{1}{2\epsilon}(X+X^\dagger)|\Psi_0\rangle,
\end{align}
and higher-order basis states are obtained recursively as
\begin{align}
    |\Psi_n\rangle = \frac{1}{2\epsilon}(X+X^\dagger)|\Psi_{n-1}\rangle.
    \label{eq:psi_recursion}
\end{align}
Up to normalization, this means $|\Psi_n\rangle \propto (X+X^\dagger)^n|\Psi_0\rangle$. In the limit $\epsilon\to0$, $(X+X^\dagger)$ approaches $2\epsilon\hat{H}$ as
\begin{align}
    ie^{-i\epsilon\hat{H}}=i(I-i\epsilon\hat{H}+\mathcal{O}(\epsilon^2)),\\
    \frac{1}{2\epsilon}(X+X^\dagger) = \hat{H}+\mathcal{O}(\epsilon^2),\label{eq:error}
\end{align}
 and thus the above recursion exactly recovers the standard Krylov subspace $\{|\Psi_0\rangle,\hat{H}|\Psi_0\rangle,\hat{H}^2|\Psi_0\rangle,\dots\}$. For any finite $\epsilon$, however, the generated subspace is a smoothly deformed version of the exact Krylov space.

It is important to emphasize that $\epsilon$ in QKUD is not a real or imaginary time-step parameter. There is no time-step tradeoff that governing basis collapse in QKUD. The $\epsilon$, instead, serves as a variational control that tunes the geometry of the Krylov basis. As $\epsilon$ deviates from zero, the basis vectors $|\Psi_n\rangle$ are gradually deformed away from the strict power-of-$\hat{H}$ form. This continuous deformation can improve numerical conditioning of the subspace without introducing the systematic errors that a discrete time-step would entail. In practice, we observe that QKUD remains stable over a broad range of $\epsilon$ values, and in the $\epsilon\to0$ limit it converges to the exact Krylov algorithm, as expected.

To evaluate eigenvalues in this variational Krylov basis, we express the required matrix elements that project the Hamiltonian on the Krylov subspace as expectation values that can be measured on the quantum processor. The $i,j$ Matrix elements for the Hamiltonian ($M_{i,j}$) and overlap matrices ($S_{i,j}$) in QKUD
 can be written as
\begin{align}
\begin{split}
    \langle\Psi_i|\hat{H}|\Psi_j\rangle
    &=\frac{1}{(2\epsilon)^{i+j}}\langle\Psi_0|(X+X^\dagger)^{i\dagger}\hat{H}(X+X^\dagger)^j|\Psi_0\rangle,\\
    \langle\Psi_i|\Psi_j\rangle
    &=\frac{1}{(2\epsilon)^{i+j}}\langle\Psi_0|(X+X^\dagger)^{i\dagger}(X+X^\dagger)^j|\Psi_0\rangle, 
\end{split}
\end{align}
which involves an expectation value evaluated on the initial state $|\Psi_0\rangle$ with unitary operations $X$ and $X^\dagger$ applied.
Using these measured matrix elements, we set up the generalized eigenvalue problem
\begin{align}\label{eq:mcsce}
    MC=SCE,
\end{align}
which is solved on a classical computer to obtain the eigenvalues $E$. 
The coefficients in $C$ define the linear combination of the $|\Psi_i\rangle$ that approximates the eigenstate. In practice, we grow the Krylov subspace incrementally one Krylov vector at a time. 
One convergence criterion is that the lowest eigenvalue has converged within a tolerance, $|E_n-E_{n-1}|<\delta$, at which point adding further basis vectors does not significantly change the result. Another possible criterion is based on linear dependence: if the newly generated $|\Psi_n\rangle$ is nearly redundant with the existing subspace, then the effective dimension of the Krylov space has been saturated. We define the effective rank of the overlap matrix $S$ as the number of eigenvalues $\lambda_i$ of $S$ that exceed a small cutoff (e.g.\ $\lambda_i>10^{-8}$); this provides a quantitative measure of the independent subspace size.

Finally, we provide two implementation proposals (provided in detail in SI) of QKUD on quantum hardware. The first is the straightforward implementation of Eq.\eqref{eq:psi_recursion} for each new Krylov vector but that would require fault-tolerant hardware. The second is a hardware-friendly implementation, that is carried out by realizing $(X+X^\dagger)$ operations by breaking down the linear combination of operators into independent expectation value measurements; for example, the first expectation value can be written as
\begin{align}
\begin{split}
&\bra{\Psi_1}H\ket{\Psi_1}=\frac{1}{(2\epsilon)^2}\bra{\Psi_0}(\hat{X}+\hat{X}^\dagger )H(\hat{X}+\hat{X}^\dagger)\ket{\Psi_0},\\
&=\frac{1}{(2\epsilon)^2}\bra{\Psi_0}\hat{X}H\hat{X}\ket{\Psi_0}
+\frac{1}{(2\epsilon)^2}\bra{\Psi_0}\hat{X}^\dagger H\hat{X}\ket{\Psi_0}\\
&+\frac{1}{(2\epsilon)^2}\bra{\Psi_0}\hat{X}H\hat{X}^\dagger \ket{\Psi_0}
+\frac{1}{(2\epsilon)^2}\bra{\Psi_0}\hat{X}^\dagger H\hat{X}^\dagger\ket{\Psi_0}.
\end{split}
\end{align}
Each of these expectation values requires prepared states of the form
$e^{\pm i\epsilon n_i H}\ket{\Psi_0}$ (forward and backward time-evolution terms),
which are as complex to prepare as in QRTE. These quantum measurements are then
followed by classical postprocessing. Although the QKUD vectors are constructed
from forward- and backward-evolved states, their final combination reconstructs
the exact action of $H$ on $\ket{\Psi_0}$ rather than a time-evolved state.
QKUD has no worse asymptotic scaling in shot count, circuit depth, or qubit
count than standard subspace-diagonalization Krylov approaches (e.g., QRTE and
related filter-diagonalization methods), which require $O(i^2N^4)$ shots
without simplifications, where $i$ and $N$ are the numbers of iterations and
qubits, respectively. In the hardware-friendly version we propose, the number
of quantum measurements is only four times that of standard QRTE while
maintaining the same gate depth. This simplicity makes QKUD a promising and
practical approach for current and future quantum devices.

\begin{figure*}[t]
  \centering
  \begin{subfigure}[b]{0.48\textwidth}
    \centering
    \includegraphics[width=\linewidth]{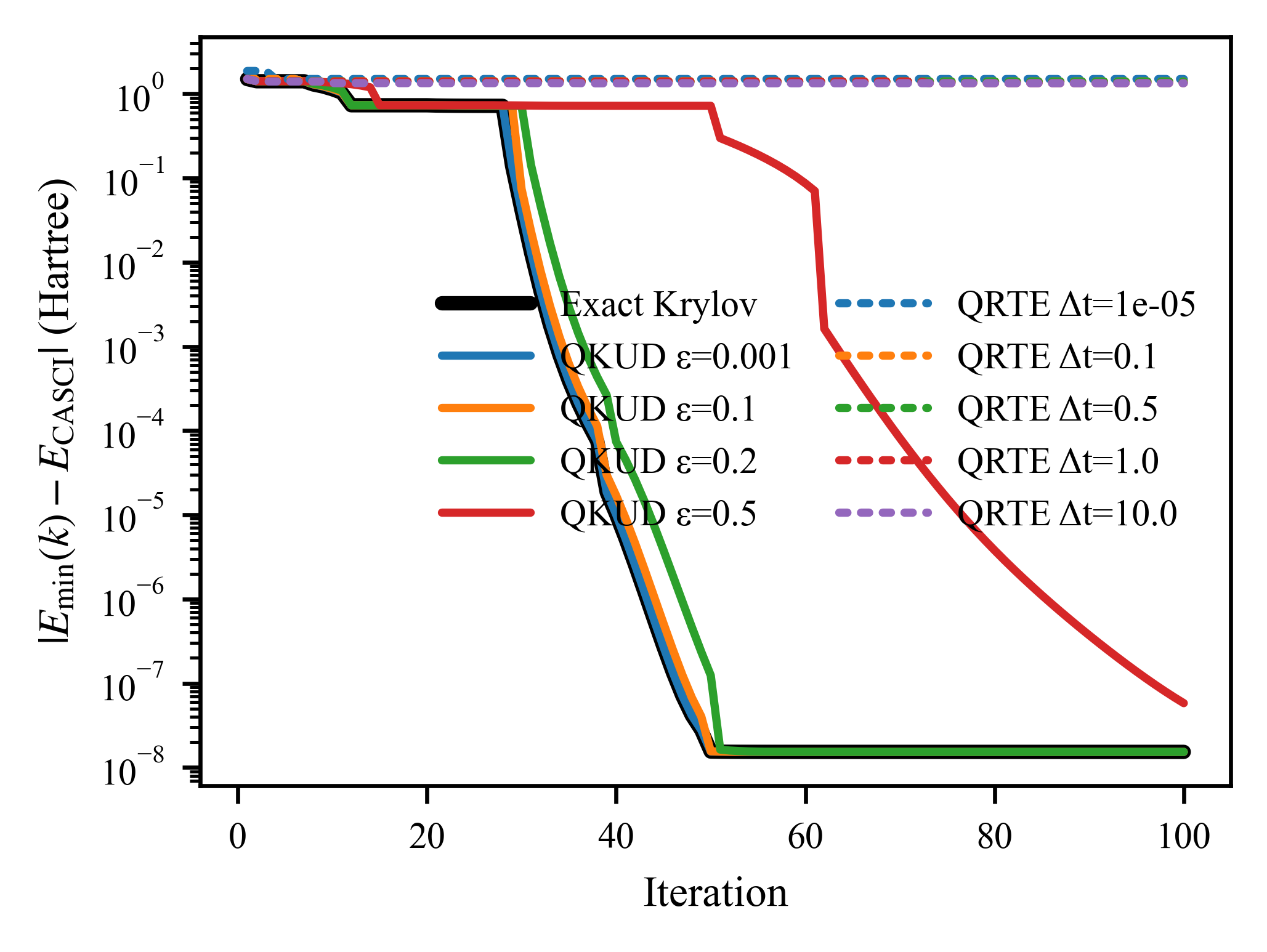}
    \caption{H$_6$ (6e,6o), $R=5.0$~\AA, STO-3G}
    \label{fig:abcd:c}
  \end{subfigure}\hfill
  \begin{subfigure}[b]{0.48\textwidth}
    \centering
    \includegraphics[width=\linewidth]{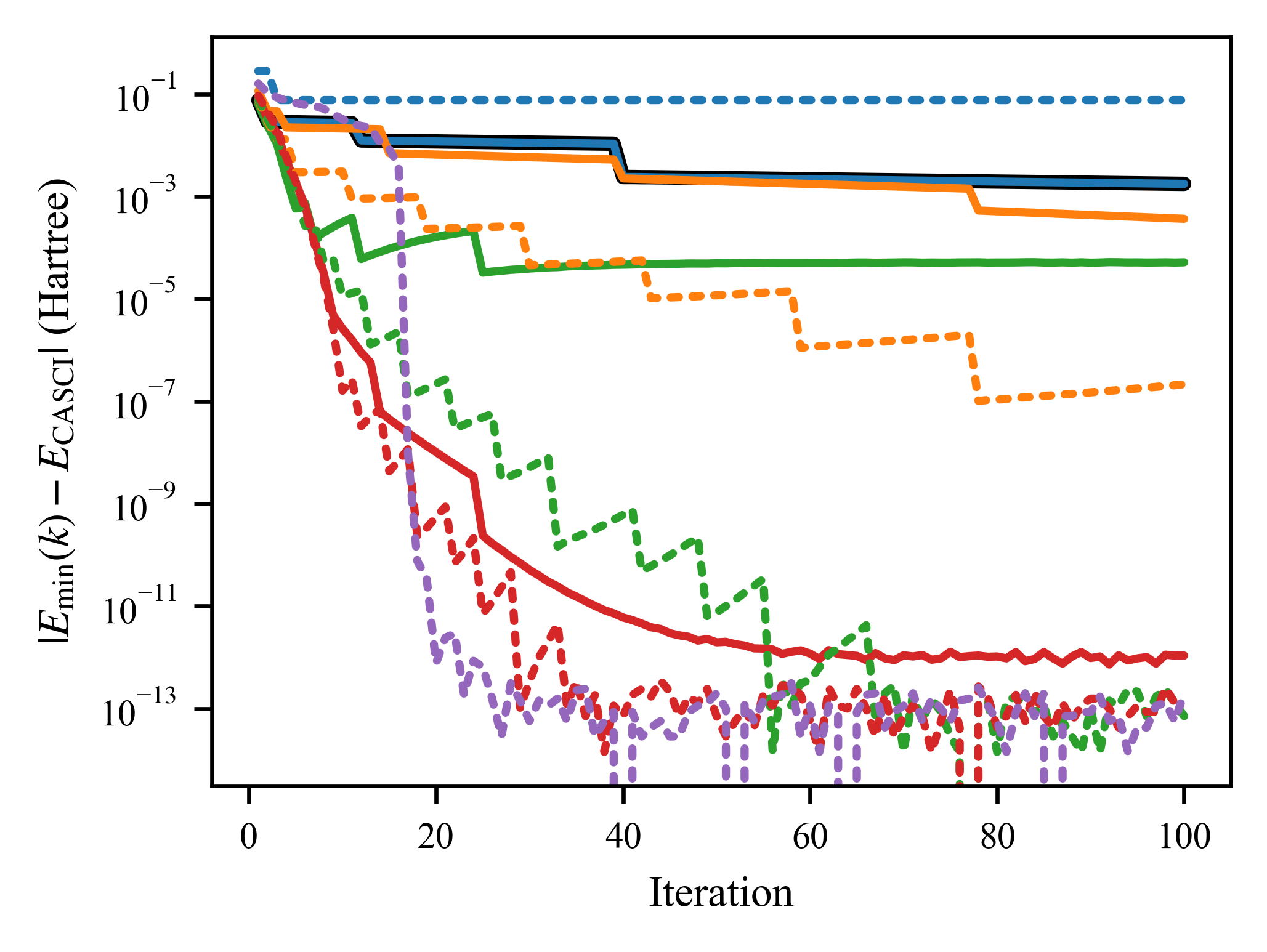}
    \caption{N$_2$ (6e,6o), $R=1.5$~\AA, STO-6G}
    \label{fig:abcd:b}
  \end{subfigure}


  \begin{subfigure}[b]{0.48\textwidth}
    \centering
    \includegraphics[width=\linewidth]{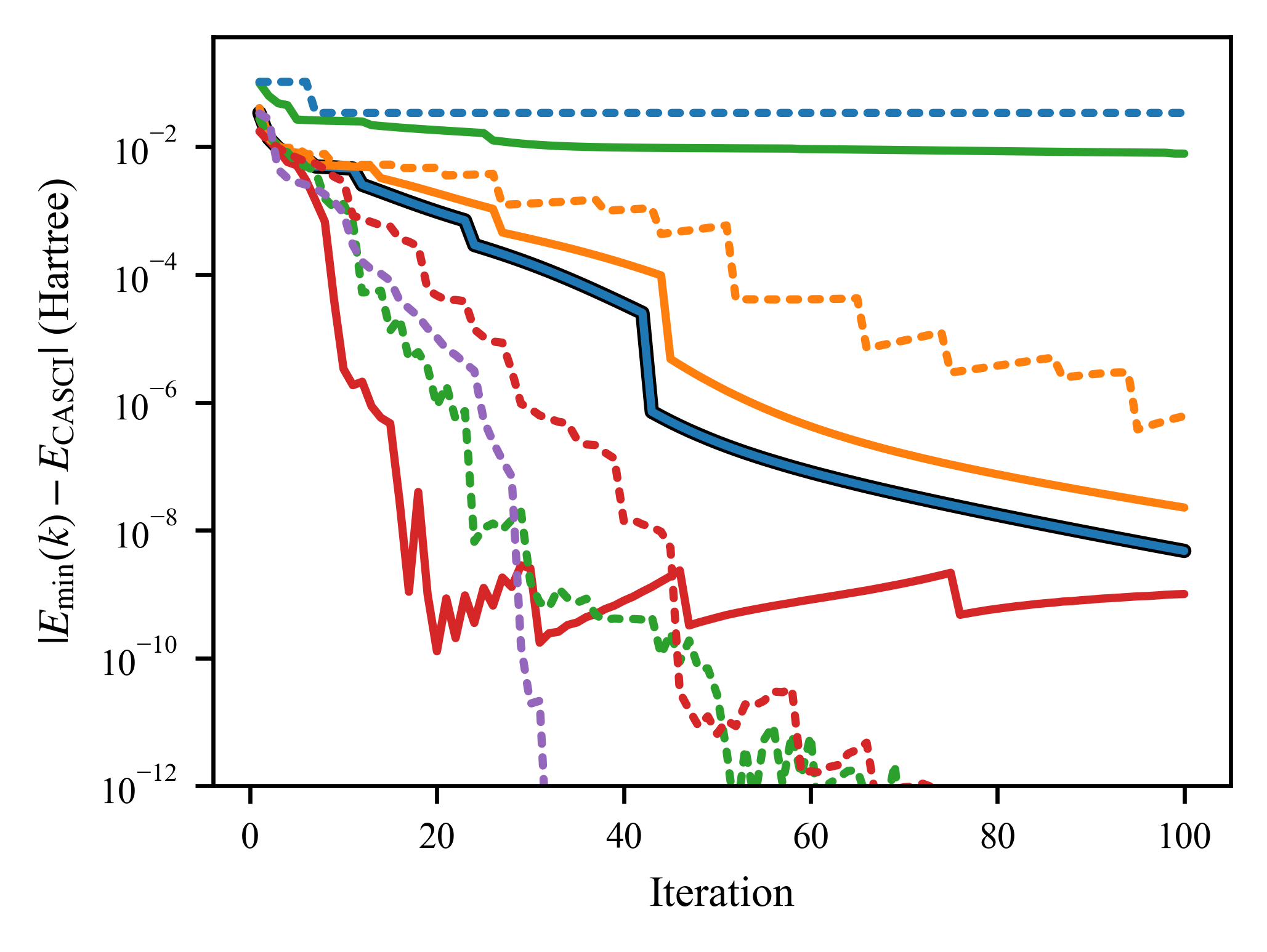}
    \caption{LiH (6e,6o), $R=3.2~\AA$, STO-3G}
    \label{fig:abcd:a}
  \end{subfigure}\hfill
  \begin{subfigure}[b]{0.48\textwidth}
    \centering
    \includegraphics[width=\linewidth]{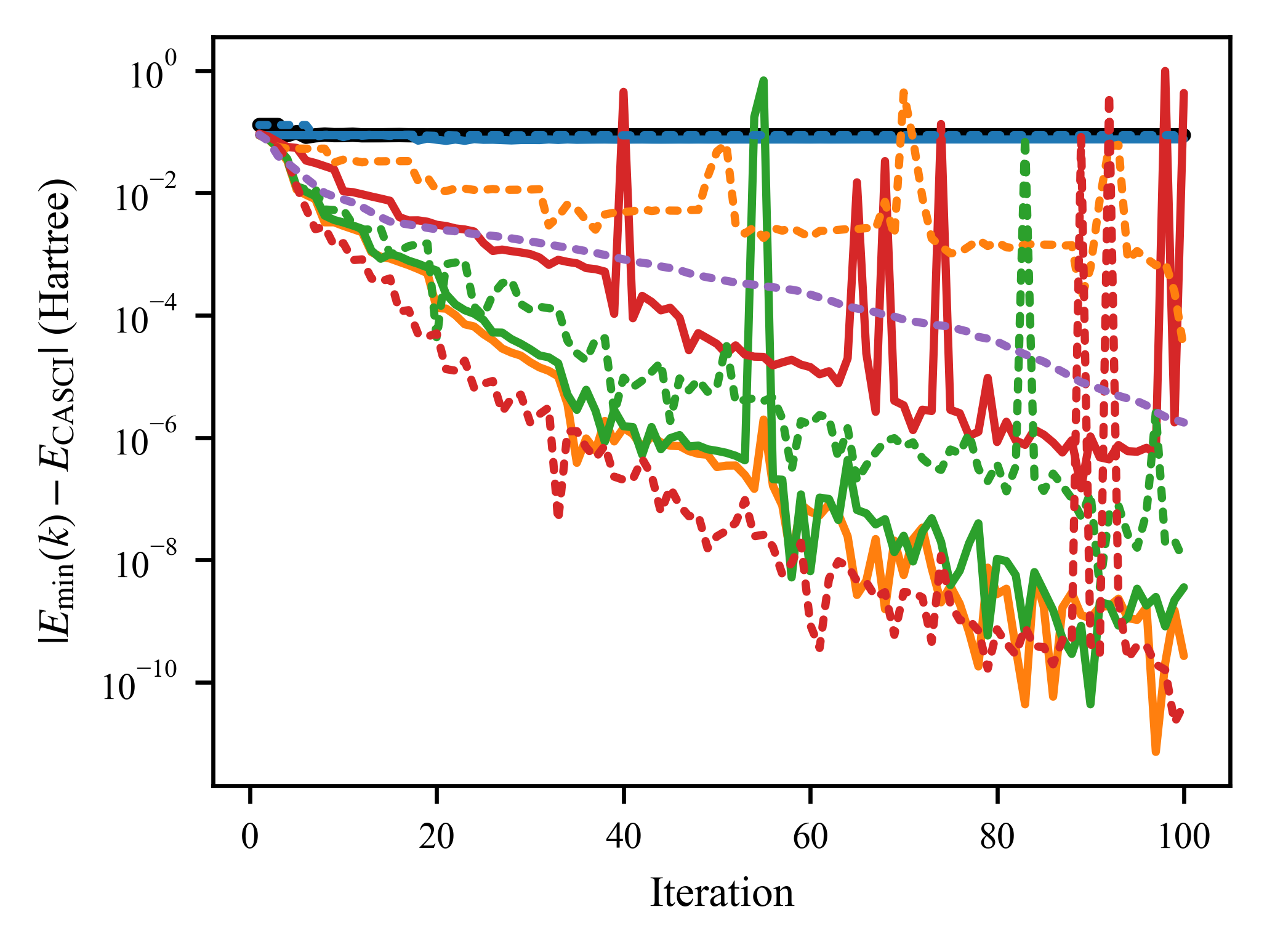}
    \caption{U$_2$ (6e,6o), $R=2.4$~\AA, ANO-RCC-MB}
    \label{fig:abcd:d}
  \end{subfigure}

  \caption{Convergence of the energy error $E_{\min}(k)-E_{\rm CASCI}$ versus iteration $k$ for
  (a) H$_6$ ($R=5.0$~\AA, STO-3G),
  (b) N$_2$ ($R=1.5$~\AA, STO-6G),
  (c) LiH ($R=3.2$~\AA, STO-3G), and
  (d) U$_2$ ($R=2.4$~\AA, ANO-RCC-MB)
  in the (6e,6o) active space. Exact Krylov is compared with QKUD and QRTE with various parameter values. Differences in convergence primarily reflect the conditioning of the generated subspace rather than time-evolution accuracy.}
  \label{fig:abcd}
\end{figure*}

\begin{figure*}[t]
  \centering
  \begin{subfigure}[b]{0.48\textwidth}
    \centering
    \includegraphics[width=\linewidth]{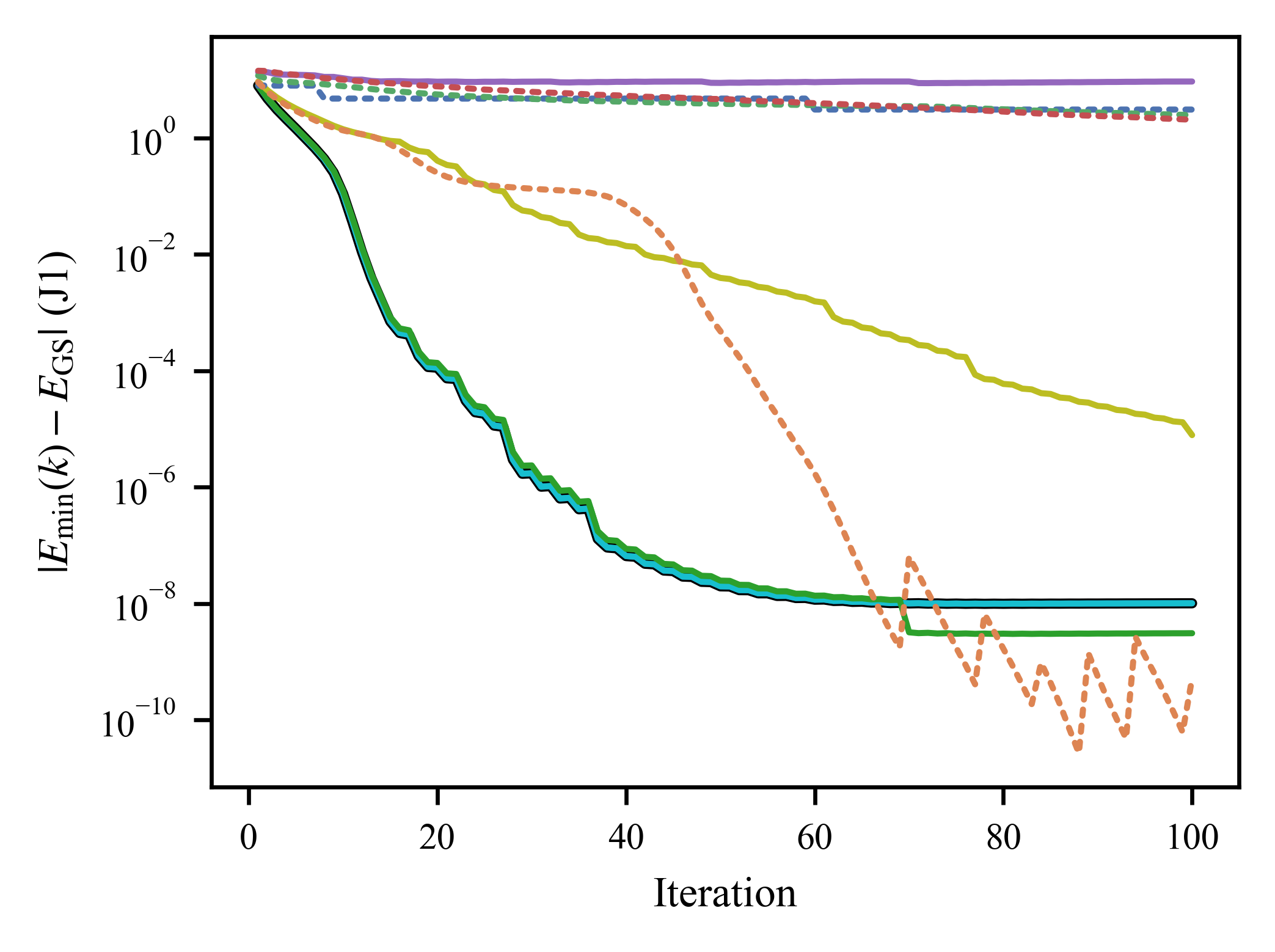}
    \caption{$4\times4$ sites}
    \label{fig:frustrated2d:c}
  \end{subfigure}\hfill
  \begin{subfigure}[b]{0.48\textwidth}
    \centering
    \includegraphics[width=\linewidth]{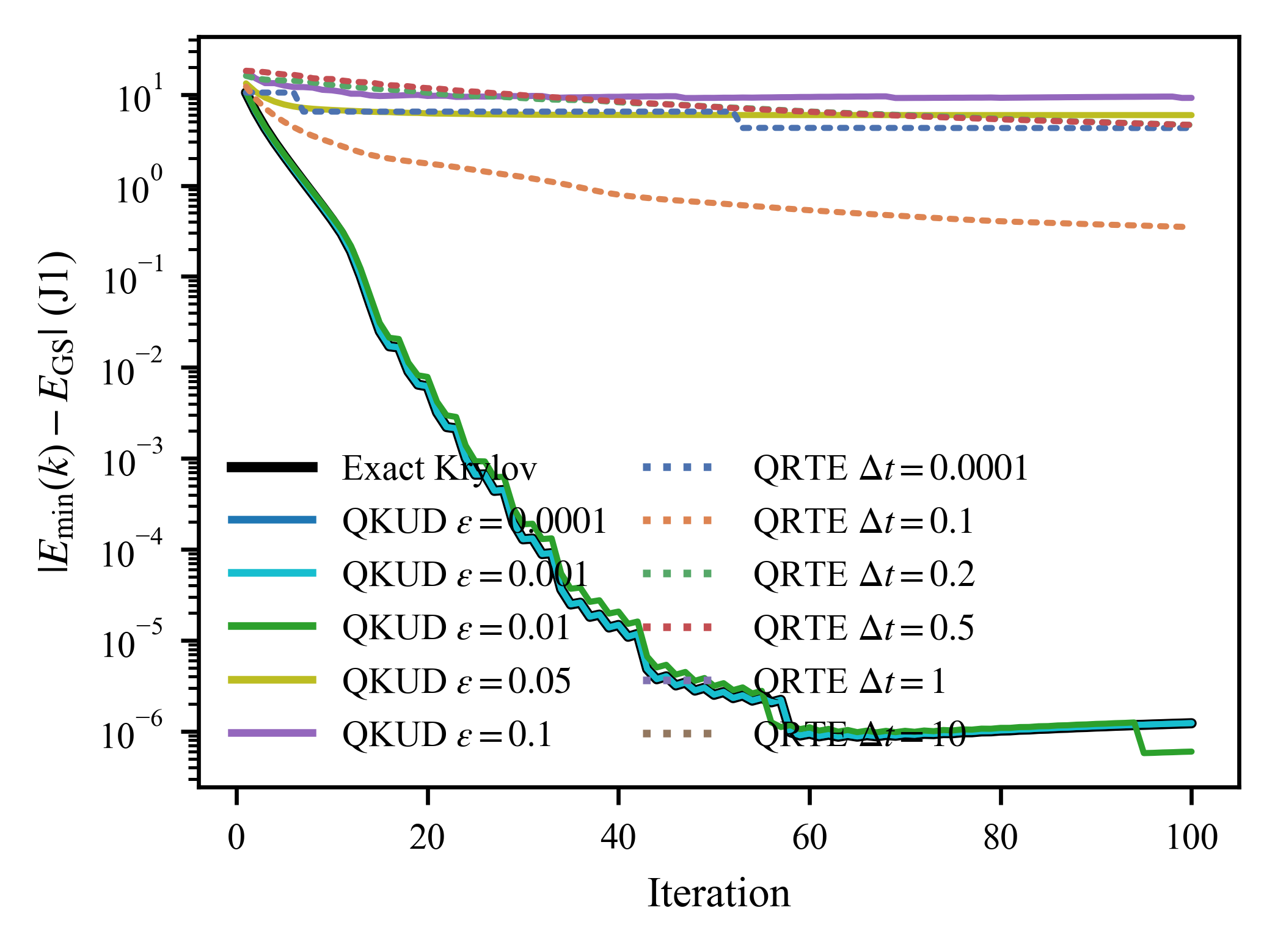}
    \caption{$4\times5$ sites}
    \label{fig:frustrated2d:d}
  \end{subfigure}

  \caption{Frustrated 2D Heisenberg model at $J_2/J_1=0.5$: energy error (relative to exact diagonalization) versus Krylov iteration $k$ for increasing system size: (a) $4\times4$, and (b) $4\times5$ sites.
  As the system size increases, a QRTE-generated Krylov basis built from fixed timesteps $\Delta t$ becomes increasingly unreliable (no single $\Delta t$ yields consistent convergence), illustrating the breakdown of a fixed-$\Delta t$ strategy at larger sizes.}
  \label{fig:frustrated2d}
\end{figure*}

\subsection{QRTE: Qauntum Real-Time Evolution}
In time-evolution-based Krylov methods, QRTE is the most popular method. In QRTE-based quantum Krylov methods and closely connected filter-diagonalization schemes ~\cite{motta2020determining,parrish2019quantum,Cohn_DF_2021}
the basis is generated by discrete real-time propagation under $\hat H$,
\begin{align}
U(\Delta t)&=e^{-i\hat H\Delta t},\\
\qquad |\Phi_n\rangle &= U(\Delta t)^n|\Psi_0\rangle,\;\; n=0,\ldots,K-1.
\end{align}
Writing $|\Psi_0\rangle=\sum_j c_j|E_j\rangle$ gives $|\Phi_n\rangle=\sum_j c_j e^{-iE_j n\Delta t}|E_j\rangle$,
so the subspace is controlled by the sampled phases $e^{-iE_j\Delta t}$. The matrices entering the
generalized eigenvalue problem ~\eqref{eq:mcsce} can be expressed as time-correlation functions,
\begin{align}
\begin{split}
S_{i,j} &= \langle \Phi_i|\Phi_j\rangle,\\
M_{i,j} &= \langle \Phi_i|\hat H|\Phi_j\rangle.
\end{split}
\end{align}
Filter diagonalization replaces $\{|\Phi_n\rangle\}$ by energy-windowed combinations
$|\chi_\mu\rangle=\sum_{n=0}^{K-1} w_{\mu n}|\Phi_n\rangle$, which induces a spectral filter
$F_\mu(E)=\sum_{n=0}^{K-1} w_{\mu n}e^{-iEn\Delta t}$ so that $|\chi_\mu\rangle=\sum_j c_j F_\mu(E_j)|E_j\rangle$.
A key practical distinction from QKUD is that $\Delta t$ simultaneously sets basis geometry and spectral sampling:
for small $\Delta t$, $U(\Delta t)\approx I-i\hat H\Delta t$ and successive $|\Phi_n\rangle$ become nearly linearly
dependent (basis collapse), while larger $\Delta t$ can reduce collapse but changes the induced span through the
$2\pi/\Delta t$ periodicity of $E\mapsto e^{-iE\Delta t}$. In contrast, QKUD tunes subspace geometry via the
deformation parameter $\epsilon$ in $(X+X^\dagger)/(2\epsilon)=\hat H+\mathcal{O}(\epsilon^2)$~\eqref{eq:error} without
introducing any physical time discretization.


\section{Results and discussion}\label{results}

The benchmarks in Figs.~\ref{fig:abcd}--\ref{fig:frustrated2d} evaluate a single practical question: under realistic iteration and measurement budgets, what controls whether a Krylov-type variational procedure continues to make progress? Across both lattice and molecular problems, the trends consistently indicate that the dominant limiter is the conditioning (linear independence) of the generated basis and the resulting stability of the generalized eigenvalue problem, rather than the nominal fidelity of the primitive used to generate trial vectors. This distinction is particularly sharp for QRTE-style constructions. While taking $\Delta t\!\to\!0$ improves dynamical faithfulness, it simultaneously drives successive time-evolved vectors toward near-linear dependence (since $e^{-i\hat H\Delta t}\approx I-i\hat H\Delta t$), causing basis collapse and ill-conditioned overlap matrices. Increasing $\Delta t$ can mitigate collapse by separating eigenphases more strongly, but it also introduces system-dependent distortions of the induced span; consequently, the performance of fixed-$\Delta t$ QRTE becomes a tuning problem whose solution need not transfer across systems. QKUD addresses the same bottleneck by operating directly at the level of Hamiltonian-power subspaces: it reduces to strict exact Krylov recursion as $\epsilon\!\to\!0$ and introduces a controlled $O(\epsilon^2)$ deformation at finite $\epsilon$~\eqref{eq:error}, so $\epsilon$ tunes subspace geometry without appealing to discretized physical time.

\begin{figure*}[t]
  \centering
  \begin{subfigure}[t]{0.48\textwidth}
    \centering
    \includegraphics[width=\linewidth]{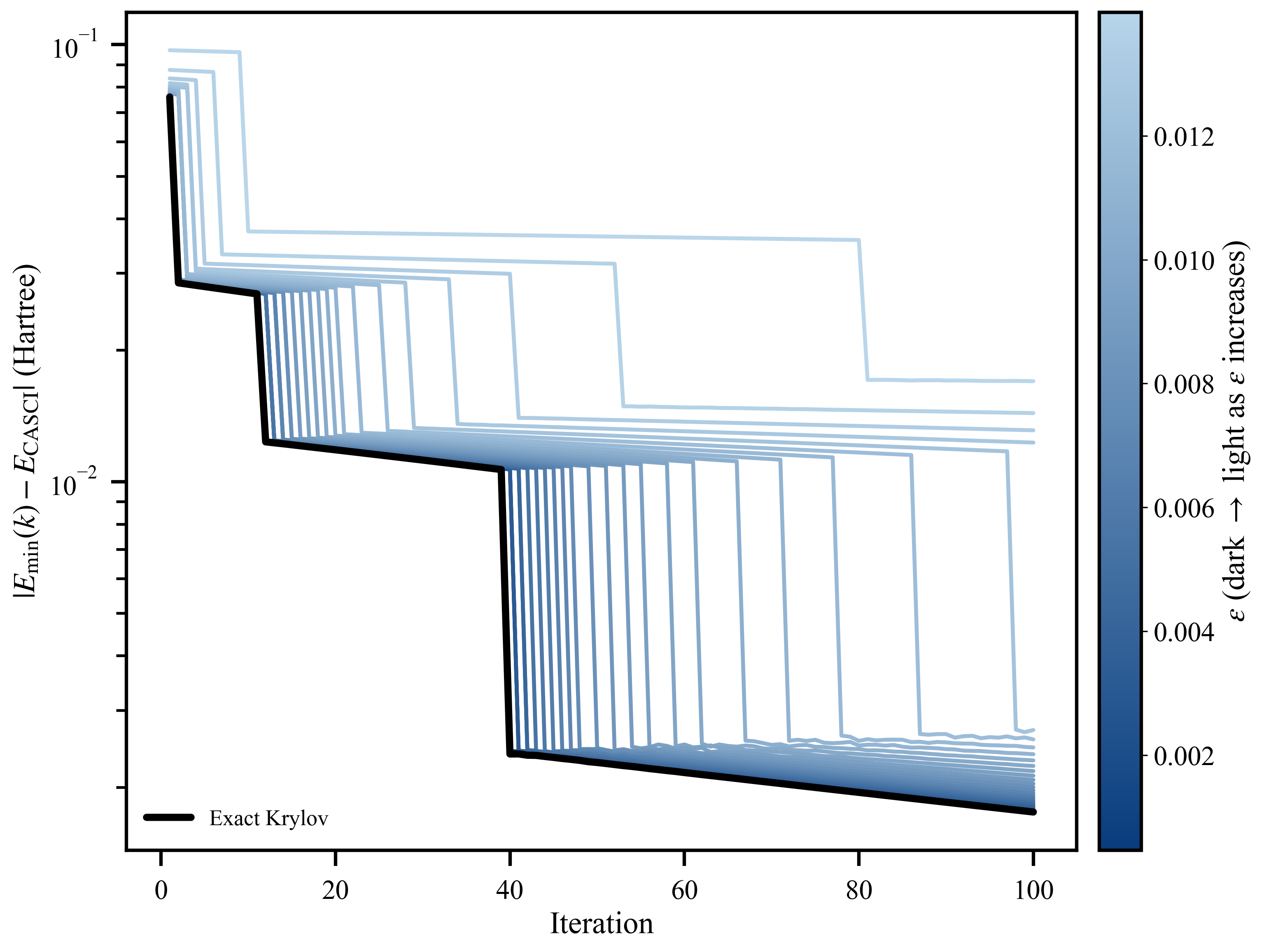}
    \caption{$\epsilon<\pi/(2\|H\|)$}
    \label{fig:n2_eps_panels:a}
  \end{subfigure}\hfill
  \begin{subfigure}[t]{0.48\textwidth}
    \centering
    \includegraphics[width=\linewidth]{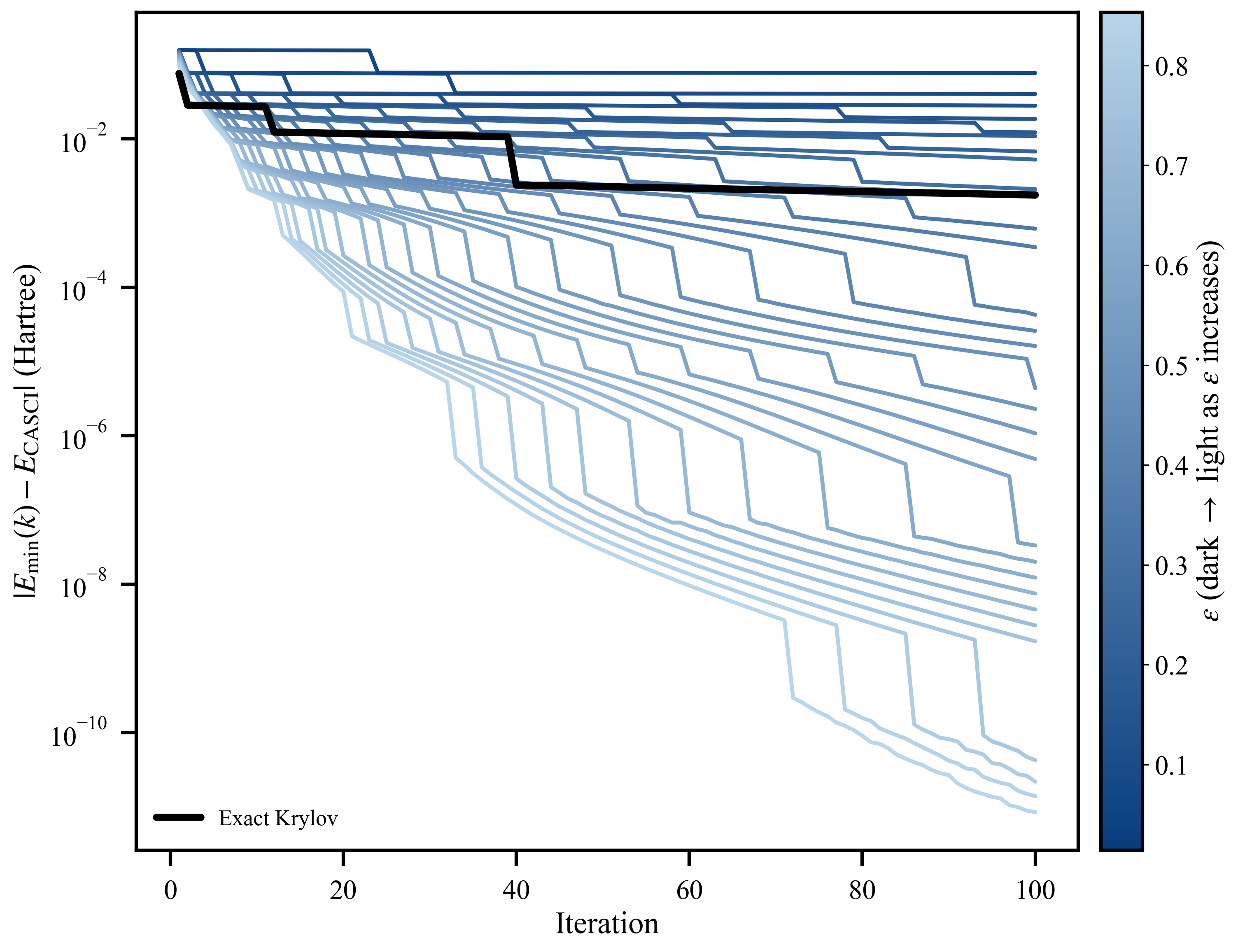}
    \caption{$\epsilon=(2n+1)\pi/(2\|H\|)$}
    \label{fig:n2_eps_panels:b}
  \end{subfigure}

  \caption{N$_2$ at $R=1.5$~\AA\ (STO-6G): QKUD behavior for two representative deformation-parameter choice, shown for (a) $\epsilon<\pi/(2\|H\|)$ and (b) $\epsilon=(2n+1)\pi/(2\|H\|)$. $\epsilon$ is chosen to be between 0 and $\frac{\pi}{2|H|}$ in (a) and chosen as the odd periods of $\frac{\pi}{2|H|}$ in (b). In both cases $\epsilon$ smoothly deforms the geometry that alows it to reach close to exact Krylov solution in (a) and very high accuracies compared with exact solver solution in (b).}
  \label{fig:n2_eps_panels}
\end{figure*}

\subsection{Varied correlation character in chemical model systems}

The molecular active-space benchmarks in Fig.~\ref{fig:abcd} are chosen to carefully span a wide range of electronic correlation regimes ranging from moderate to challenging in the active space size~\cite{Sundar2026ChemicallyDecisive}. It shows the following behavior in chemically relevant settings, with errors reported relative to CASCI. First, QRTE remains strongly $\Delta t$ dependent and can fail to provide a usable basis within a practical iteration window. This is most apparent in H$_6$ (Fig.~\ref{fig:abcd}), where, across the explored timesteps, QRTE does not consistently access the low-error regime within the iteration budget: small $\Delta t$ values plateau early in a manner consistent with collapse, while larger $\Delta t$ values do not reliably recover monotonic convergence. QKUD, by contrast, exhibits stable convergence in this case and closely follows Exact Krylov at small $\epsilon$, consistent with its $\epsilon\!\to\!0$ limit. Second, several molecular panels highlight a complementary regime in which strict (non-orthogonalized) Krylov recursion itself saturates early within the tested iteration window. In such cases, taking $\epsilon$ very small forces QKUD to reproduce that same saturation, whereas moderate $\epsilon$ values deliberately reshape the span enough to restore variational improvement. This behavior is clearly reflected in instances where the Exact Krylov reference stalls at a relatively large residual error while larger $\epsilon$ choices (and, in QRTE, certain larger timesteps) achieve additional orders-of-magnitude improvement. The natural interpretation is that, once the overlap spectrum develops extremely small eigenvalues, the procedure stops adding genuinely new independent information; controlled distortion can reweight the span so that the generalized eigenproblem remains informative and the Ritz value continues to improve. Across the molecular benchmarks studied here, QKUD is consistent and reaches chemical accuracy in every case.

\subsection{Generality: frustrated \texorpdfstring{$J_1$--$J_2$}{J1-J2} Heisenberg Hamiltonian on a Square Lattice}

In Fig. \ref{fig:frustrated2d}, we study a spin-$\tfrac12$ model on an $L_x \times L_y$ square lattice with site coordinates
$(x,y)$, where $x=0,\dots,L_x-1$ and $y=0,\dots,L_y-1$.  
The implementation maps each site to a qubit.
The Hamiltonian is
\begin{align}
\begin{split}
\hat H
&= J_1 \sum_{\langle i,j\rangle}
\left(
\hat X_i \hat X_j + \hat Y_i \hat Y_j + \hat Z_i \hat Z_j
\right)\\
&+ J_2 \sum_{\langle\!\langle i,j\rangle\!\rangle}
\left(
\hat X_i \hat X_j + \hat Y_i \hat Y_j + \hat Z_i \hat Z_j
\right),
\label{eq:j1j2_pauli}
\end{split}
\end{align}
where $\hat X,\hat Y,\hat Z$ are Pauli operators
Nearest-neighbor bonds $\langle i,j\rangle$ are generated along the horizontal and vertical directions:
\begin{equation}
(x,y)\leftrightarrow(x+1,y), \qquad (x,y)\leftrightarrow(x,y+1),
\end{equation}
while next-nearest-neighbor (frustrating) bonds $\langle\!\langle i,j\rangle\!\rangle$ are taken along both diagonals:
\begin{equation}
(x,y)\leftrightarrow(x+1,y+1), \qquad (x,y)\leftrightarrow(x+1,y-1).
\end{equation}
Each bond is included once. Open boundaries with only in-range pairs are kept.
In the numerical analysis, couplings are parameterized as
\begin{equation}
J_1 = 1, \qquad J_2 = r\,J_1,
\end{equation}
with $r \equiv J_2/J_1$ controlling frustration that we set to 0.5 which places the model in the frustrated regime.

Figure~\ref{fig:frustrated2d} provides a stringent scalability test reporting the ground-state energy error relative to exact diagonalization as a function of Krylov iteration for increasing lattice size of the frustrated 2D model. For the smaller lattices, QRTE can yield strong convergence for $\Delta t=0.1$, but this behavior is not scalable: as the lattice grows from $4\times 4$ to $4\times 5$, fixed-$\Delta t$ QRTE becomes increasingly unreliable, with different timesteps exhibiting stagnation or inconsistent late-iteration behavior and no single $\Delta t$ producing uniform convergence across sizes. This is consistent with the underlying spectral picture: as the many-body spectrum becomes denser and more structured with system size, the window of timesteps that avoids both (i) collapse from near-linear dependence and (ii) unhelpful distortion narrows and shifts in an instance-dependent manner. In contrast, QKUD closely tracks the Exact Krylov reference whenever strict recursion remains well conditioned, and it maintains stable convergence across the full size sweep for small-to-moderate $\epsilon$. The key point is not that time evolution becomes generically “less accurate” with size, but that fixed-$\Delta t$ sampling becomes a progressively more brittle method to generate a well-conditioned, variationally useful basis as many-body complexity increases. There are some many-body Hamiltonian cases where even QKUD can exhibit linear dependence, especially when the exact Krylov saturates quickly; these cases may be assisted by using a shifted Hamiltonian to reduce its norm or combining QKUD with a multireference starting state, for example as proposed in MRSQK~\cite{stair2020mrqk}.

Taken together, these results support a unified operating principle for practical quantum Krylov algorithms. When strict Krylov recursion is a good solution (i.e., the generated basis remains well conditioned over the relevant iteration range), QKUD at small $\epsilon$ is essentially indistinguishable from Exact Krylov, as expected. When strict recursion saturates within realistic iteration limits, finite $\epsilon$ provides a systematic and predictable knob to deform the basis and recover progress. QRTE, in contrast, achieves collapse-avoidance and distortion only indirectly through a fixed timestep, making its performance intrinsically $\Delta t$ dependent: for some instances it may be impossible to find any fixed $\Delta t$ that works within the available budget (as in H$_6$), and even when convergence is possible, identifying a suitable timestep becomes increasingly difficult as system size and spectral complexity grow (as in the frustrated 2D scaling). The broader implication is that scalable Krylov strategies should prioritize explicit control of overlap conditioning and subspace geometry over reliance on delicate, instance-specific timestep selection in real-time evolution constructions.

\begin{figure*}[t]
  \centering
  \begin{subfigure}[t]{0.48\textwidth}
    \centering
    \includegraphics[width=\linewidth]{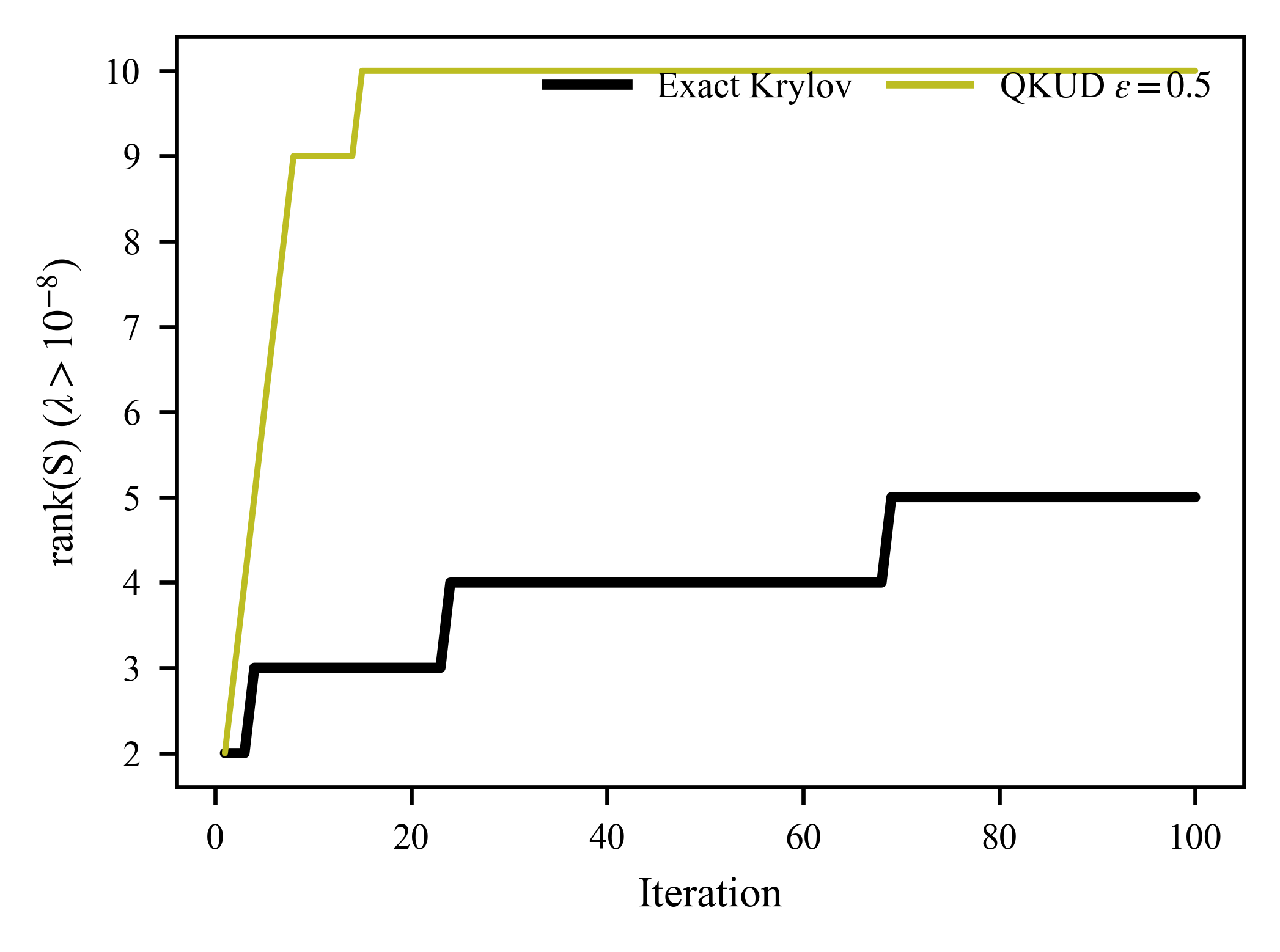}
    \caption{Effective rank of the overlap matrix $S$ vs.\ iteration $k$ (counting eigenvalues $\lambda(S)>10^{-8}$).}
    \label{fig:n2_rank_cond:a}
  \end{subfigure}\hfill
  \begin{subfigure}[t]{0.48\textwidth}
    \centering
    \includegraphics[width=\linewidth]{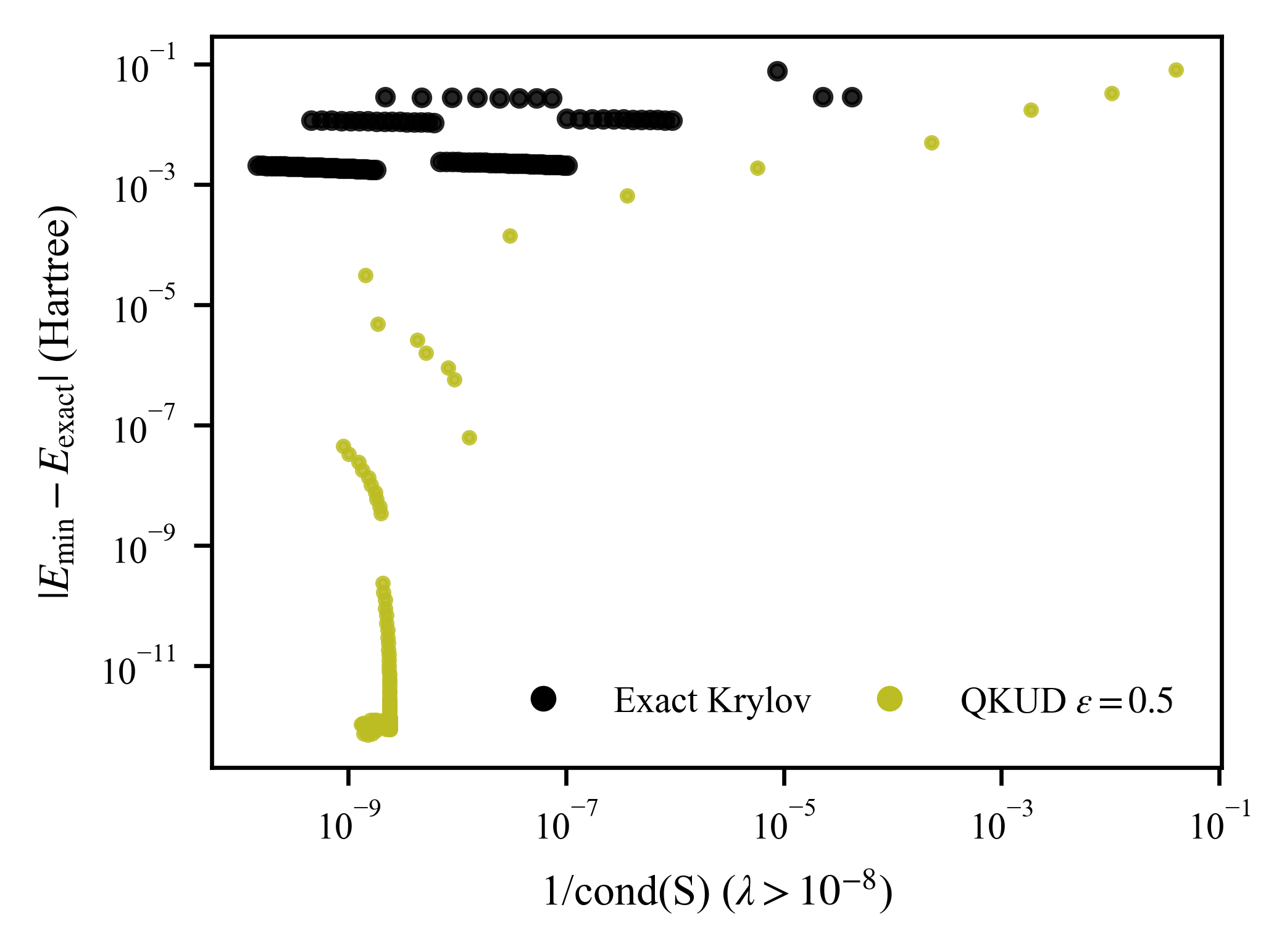}
    \caption{Energy error $\left|E_{\min}(k)-E_{\rm CASCI}\right|$ vs.\ inverse condition number $1/\mathrm{cond}(S)$.}
    \label{fig:n2_rank_cond:b}
  \end{subfigure}

  \caption{Subspace-geometry diagnostics for N$_2$ at $R=1.5$~\AA{} (STO-6G, 6e,6o).
  (a) Effective rank of the Krylov overlap matrix $S$ as the subspace grows.
  (b) Energy error plotted against $1/\mathrm{cond}(S)=\sigma_{\min}(S)/\sigma_{\max}(S)$.
  Improved energy in QKUD correlate with rank growth and maintaining a well-conditioned overlap matrix.}
  \label{fig:n2_rank_cond}
\end{figure*}

\subsection{Role of deformation parameter $\epsilon$}
QKUD is built from the Hermitian generator
\begin{equation}
A(\epsilon)\equiv \frac{\hat X+\hat X^\dagger}{2\epsilon}=\frac{\sin(\epsilon \hat H)}{\epsilon},
\end{equation}
which reduces to Hamiltonian-power Krylov in the small-$\epsilon$ limit,
\begin{equation}
\frac{\sin(\epsilon \hat H)}{\epsilon}=\hat H+\mathcal{O}(\epsilon^2).
\end{equation}
Finite $\epsilon$ therefore does not represent a time step; it smoothly deforms the Krylov span by a controlled spectral reweighting. In the eigenbasis $\hat H|E\rangle=E|E\rangle$,
\begin{equation}
A(\epsilon)|E\rangle = f_\epsilon(E)\,|E\rangle,\qquad
f_\epsilon(E)=\frac{\sin(\epsilon E)}{\epsilon}=E\,\mathrm{sinc}(\epsilon E),
\end{equation}
so increasing $\epsilon$ damps large-$|E|$ components through the $\mathrm{sinc}$ factor and can improve the conditioning of the overlap matrix without sacrificing Hermiticity.

A convenient operating window is set by the Hamiltonian norm $||H||\equiv\max |E|$. Different strategies for estimating $||H||$ such as through use of Jordon-Wignerized Pauli Hamiltonian. When $\epsilon<\pi/(2||H||)$, the map $E\mapsto \sin(\epsilon E)$ is monotone across the full spectrum, so QKUD remains strongly Krylov-like while still benefitting from mild spectral damping. This behavior is consistent with our N$_2$ results in Fig.~\ref{fig:n2_eps_panels:a}, where trajectories for $\epsilon\in(0,\pi/(2||H||))$ track the exact-Krylov trend and show stable, predictable convergence.

Beyond this window, the spectral reweighting becomes more aggressive and can be exploited deliberately to mitigate collinearity of Kryov vectors and generalized-eigenproblem ill-conditioning that may cause the exact-Krylov baseline to stall at larger subspace sizes. A particularly robust practical choice is a discrete schedule
\begin{equation}
\epsilon_n = \frac{n\pi}{||H||} \text{ or } \frac{(2n+1)\pi}{2||H||},
\end{equation}
which periodically suppresses extremal spectral contributions and refreshes the effective geometry of the generated vectors. Empirically, this schedule yields a more consistent convergence story on N$_2$ (Fig.~\ref{fig:n2_eps_panels}), with many runs continuing to improve even in regimes where the exact-Krylov curve saturates. For this reason, we recommend $\epsilon_n=n\pi/||H||$ or $\frac{(2n+1)\pi}{2||H||}$ as default, especially as a fallback strategy when overlap conditioning limits further progress in the undeformed Krylov recursion. 

Relative to QRTE, the key distinction is that QRTE tunes subspace geometry indirectly through the unitary step $U(\Delta t)=e^{-i\hat H\Delta t}$, where too small $\Delta t$ hints at near-linear dependence ($U(\Delta t)\approx I-i\hat H\Delta t$) and larger $\Delta t$ introduces phase wrapping through the periodic map $E\mapsto e^{-iE\Delta t}$. In contrast, QKUD uses the Hermitian, analytic transform $\sin(\epsilon\hat H)/\epsilon$, so $\epsilon$ acts as a direct and controllable deformation knob, with a natural monotone regime and a simple discrete schedule that can be used to improve robustness when conditioning becomes the limiting factor.

\subsection{Geometry analysis}
Figure~\ref{fig:n2_rank_cond:a} and Fig.~\ref{fig:n2_rank_cond:b} directly diagnose how subspace geometry limits convergence for N$_2$ at $R=1.5$~\AA\ and how QKUD alleviates this bottleneck. In Fig.~\ref{fig:n2_rank_cond:b}, we plot the achieved energy error $|E_{\min}-E_{\mathrm{exact}}|$ versus $1/\mathrm{cond}(S)$, where $\mathrm{cond}(S)$ is computed after discarding overlap eigenvalues below $10^{-8}$; moving left corresponds to a more ill conditioned overlap and a harder generalized eigenproblem. Exact Krylov (black) clusters at relatively large errors even deep in the ill conditioned regime, indicating that the nominal increase in iteration does not translate into additional usable directions in the span and the solve effectively stalls. In contrast, QKUD at fixed $\epsilon=0.5$ (yellow) accesses orders-of-magnitude smaller errors while operating at comparable values of $1/\mathrm{cond}(S)$, consistent with the deformation producing vectors that remain informative for the target eigenspace rather than collapsing into near-collinearity. This interpretation is corroborated by Fig.~\ref{fig:n2_rank_cond:a}, which reports the effective rank of $S$ (counting eigenvalues $\lambda>10^{-8}$): QKUD rapidly grows to near full rank (approaching the intended subspace dimension) within the first few tens of iterations, whereas exact Krylov remains rank limited throughout. Together, these results show that QKUD improves the usable dimensionality and stability of the subspace, enabling continued variational improvement precisely in regimes where overlap ill conditioning constrains the undeformed Krylov recursion.

\section{Conclusions}\label{conclusions}

We introduced Quantum Krylov using Unitary Decomposition (QKUD), a strategy for constructing quantum Krylov subspaces without explicit time evolution by mapping Hamiltonian powers through a unitary decomposition. QKUD is formally exact in the limit $\epsilon\to 0$, recovering strict Krylov recursion, while finite $\epsilon$ provides a controlled deformation that tunes subspace geometry rather than discretizing physical time. This separation removes timestep-selection ambiguity inherent to real-time-evolution Krylov methods and mitigates basis collapse when successive time-evolved vectors become nearly linearly dependent.

Across benchmarks spanning weakly correlated, near-degenerate, and strongly correlated regimes, QKUD reproduces exact Krylov convergence when the recursion remains well conditioned and restores convergence when strict Krylov and time-evolution-generated bases stagnate within practical iteration budgets. Geometry diagnostics show that energy improvement tracks overlap-matrix stability, including maintaining nonvanishing $\sigma_{\min}$ and avoiding large $\mathrm{cond}(S)$, more reliably than it tracks rank growth alone. This identifies overlap conditioning, not time-evolution fidelity, as the relevant bottleneck for Krylov performance on quantum hardware. Classically, basis collapse is addressed by orthogonalization or restarts; however, full re-orthonormalization of quantum states is prohibitively costly. QKUD instead improves robustness by modifying basis generation to avoid linear dependence.

For near-term and early fault-tolerant quantum simulation, these results suggest that Krylov algorithm design should prioritize control of Krylov-subspace geometry and conditioning over pursuing ever more accurate coherent time evolution. The stagnation of exact Krylov observed for N$_2$ and U$_2$ illustrates that this limitation is most severe in strongly correlated settings, and the same mechanism is expected to arise broadly in interacting many-body systems. QKUD not only provides a route to exact Krylov simulation in principle, but also offers a tunable, analytic handle for shaping the Krylov basis under realistic resource constraints, complementary to asymptotically exact approaches. More broadly, treating subspace geometry as a controllable resource suggests new directions for robust Krylov-based simulation and related iterative quantum algorithms.

\section{Acknowledgments}\label{acknowledgements}
AA acknowledges NSF awards 2427046 and 2429752 for support. AA thanks Anthony Schlimgen and Kade Head-Marsden for helpful discussions, and Ed Barnes for helpful feedback on this project. AA thanks Tushar Kumar Jain, Nick Mayhall, and Vibin Abraham for their useful comments. AA also acknowledges the UND Computational Research Center for computing resources.
\appendix


\bibliography{main}

@article{fischer2025large,
  title={Large-scale implementation of quantum subspace expansion with classical shadows},
  author={Fischer, Laurin E and Bultrini, Daniel and Tavernelli, Ivano and Tacchino, Francesco},
  journal={arXiv preprint arXiv:2510.25640},
  year={2025}
}

@article{yoshida2025auxiliary,
  title={Auxiliary-field quantum Monte Carlo method with quantum selected configuration interaction},
  author={Yoshida, Yuichiro and Erhart, Luca and Murokoshi, Takuma and Nakagawa, Rika and Mori, Chihiro and Miyanaga, Takafumi and Mori, Toshio and Mizukami, Wataru},
  journal={arXiv preprint arXiv:2502.21081},
  year={2025}
}

@article{fedorov2022vqe,
  title={VQE method: A short survey and recent developments},
  author={Fedorov, Dmitry A and Peng, Bo and Govind, Niranjan and Alexeev, Yuri},
  journal={Mat. Theory},
  volume={6},
  number={1},
  pages={1--21},
  year={2022},
  publisher={SpringerOpen}
}

@article{qdavidson,
  title={Quantum Davidson algorithm for excited states},
  author={Tkachenko, Nikolay V and Cincio, Lukasz and Boldyrev, Alexander I and Tretiak, Sergei and Dub, Pavel A and Zhang, Yu},
  journal={Quantum Science and Technology},
  volume={9},
  number={3},
  pages={035012},
  year={2024},
  publisher={IOP Publishing}
}

@article{stair2020mrqk,
author = {Stair, Nicholas H. and Huang, Renke and Evangelista, Francesco A.},
title = {A Multireference Quantum Krylov Algorithm for Strongly Correlated Electrons},
journal = {J. Chem. Theory Comput.},
volume = {16},
number = {4},
pages = {2236-2245},
year = {2020},
doi = {10.1021/acs.jctc.9b01125}
}

@article{schlimgen2021quantum,
  title={Quantum simulation of open quantum systems using a unitary decomposition of operators},
  author={Schlimgen, Anthony W and Head-Marsden, Kade and Sager, LeeAnn M and Narang, Prineha and Mazziotti, David A},
  journal = {Phys. Rev. Lett.},
  volume={127},
  number={27},
  pages={270503},
  year={2021},
  publisher={APS}
}

@article{Meitei2020,
	Author = {Meitei, Oinam Romesh and Gard, Bryan T. and Barron, George S. and Pappas, David P. and Economou, Sophia E. and Barnes, Edwin and Mayhall, Nicholas J.},
	Da = {2021/10/27},
	Doi = {10.1038/s41534-021-00493-0},
	Id = {Meitei2021},
	Isbn = {2056-6387},
	journal = {Quantum},
	Number = {1},
	Pages = {155},
	Title = {Gate-free state preparation for fast variational quantum eigensolver simulations},
	Ty = {JOUR},
	Url = {https://doi.org/10.1038/s41534-021-00493-0},
	Volume = {7},
	Year = {2021}}

@article{tilly2021variational,
  title={The Variational Quantum Eigensolver: a review of methods and best practices},
  author={Tilly, Jules and Chen, Hongxiang and Cao, Shuxiang and Picozzi, Dario and Setia, Kanav and Li, Ying and Grant, Edward and Wossnig, Leonard and Rungger, Ivan and Booth, George H and others},
  journal = {arXiv},
  year={2021}
}

@article{mcardle2020quantum,
  title={Quantum Comput. Chem.},
  author={McArdle, Sam and Endo, Suguru and Aspuru-Guzik, Al{\'a}n and Benjamin, Simon C and Yuan, Xiao},
  journal={Rev. Modern Phys.},
  volume={92},
  number={1},
  pages={015003},
  year={2020},
  publisher={APS}
}

@article{cerezo2021variational,
  title={Variational quantum algorithms},
  author={Cerezo, Marco and Arrasmith, Andrew and Babbush, Ryan and Benjamin, Simon C and Endo, Suguru and Fujii, Keisuke and McClean, Jarrod R and Mitarai, Kosuke and Yuan, Xiao and Cincio, Lukasz and others},
  journal={Nat. Rev. Phys.},
  pages={1--20},
  year={2021},
  publisher={Nature Publishing Group}
}

@article{peruzzo2014variational,
  title={A variational eigenvalue solver on a photonic quantum processor},
  author={Peruzzo, Alberto and McClean, Jarrod and Shadbolt, Peter and Yung, Man-Hong and Zhou, Xiao-Qi and Love, Peter J and Aspuru-Guzik, Al{\'a}n and O’brien, Jeremy L},
  journal = {Nat. Commun.},
  volume={5},
  number={1},
  pages={1--7},
  year={2014},
  publisher={Nature Publishing Group}
}

@article{kandala2017hardware,
  title={Hardware-efficient variational quantum eigensolver for small molecules and quantum magnets},
  author={Kandala, Abhinav and Mezzacapo, Antonio and Temme, Kristan and Takita, Maika and Brink, Markus and Chow, Jerry M and Gambetta, Jay M},
  journal={Nature},
  volume={549},
  number={7671},
  pages={242--246},
  year={2017},
  publisher={Nature Publishing Group}
}

@article{magann2021pulses,
  title={From pulses to circuits and back again: A quantum optimal control perspective on variational quantum algorithms},
  author={Magann, Alicia B and Arenz, Christian and Grace, Matthew D and Ho, Tak-San and Kosut, Robert L and McClean, Jarrod R and Rabitz, Herschel A and Sarovar, Mohan},
  journal = {Quantum},
  volume={2},
  number={1},
  pages={010101},
  year={2021},
  publisher={APS}
}

@article{grimsley2019adaptive,
  title={An adaptive variational algorithm for exact molecular simulations on a quantum computer},
  author={Grimsley, Harper R and Economou, Sophia E and Barnes, Edwin and Mayhall, Nicholas J},
  journal = {Nat. Commun.},
  volume={10},
  number={1},
  pages={1--9},
  year={2019},
  publisher={Nature Publishing Group}
}

@article{cao2019quantum,
  title={Quantum chemistry in the age of quantum computing},
  author={Cao, Yudong and Romero, Jonathan and Olson, Jonathan P and Degroote, Matthias and Johnson, Peter D and Kieferov{\'a}, M{\'a}ria and Kivlichan, Ian D and Menke, Tim and Peropadre, Borja and Sawaya, Nicolas PD and others},
  journal = {Chem. Rev.},
  volume={119},
  number={19},
  pages={10856--10915},
  year={2019},
  publisher={ACS Publications}
}

@article{motta2020determining,
  title={Determining eigenstates and thermal states on a quantum computer using quantum imaginary time evolution},
  author={Motta, Mario and Sun, Chong and Tan, Adrian TK and O’Rourke, Matthew J and Ye, Erika and Minnich, Austin J and Brand{\~a}o, Fernando GSL and Chan, Garnet Kin-Lic},
  journal={Nat. Phys.},
  volume={16},
  number={2},
  pages={205--210},
  year={2020},
  publisher={Nature Publishing Group}
}

@article{stair2020multireference,
  title={A multireference quantum krylov algorithm for strongly correlated electrons},
  author={Stair, Nicholas H and Huang, Renke and Evangelista, Francesco A},
  journal = {J. Chem. Theory Comput.},
  volume={16},
  number={4},
  pages={2236--2245},
  year={2020},
  publisher={ACS Publications}
}

@article{yu2025quantum,
  title={Quantum-centric algorithm for sample-based krylov diagonalization},
  author={Yu, Jeffery and Moreno, Javier Robledo and Iosue, Joseph T and Bertels, Luke and Claudino, Daniel and Fuller, Bryce and Groszkowski, Peter and Humble, Travis S and Jurcevic, Petar and Kirby, William and others},
  journal = {arXiv},
  year={2025}
}

@article{asthana2022minimizing,
  title={Minimizing state preparation times in pulse-level variational molecular simulations},
  author={Asthana, Ayush and Liu, Chenxu and Meitei, Oinam Romesh and Economou, Sophia E and Barnes, Edwin and Mayhall, Nicholas J},
  journal = {arXiv},
  year={2022}
}

@article{Cohn_DF_2021,
  title = {Quantum Filter Diagonalization with Compressed Double-Factorized Hamiltonians},
  author = {Cohn, Jeffrey and Motta, Mario and Parrish, Robert M.},
  journal = {Quantum},
  volume = {2},
  issue = {4},
  pages = {040352},
  numpages = {19},
  year = {2021},
  month = {Dec},
  publisher = {American Physical Society},
  doi = {10.1103/PRXQuantum.2.040352},
  url = {https://link.aps.org/doi/10.1103/PRXQuantum.2.040352}
}

@article{danilov2025enhancing,
  title={Enhancing the accuracy and efficiency of sample-based quantum diagonalization with phaseless auxiliary-field quantum Monte Carlo},
  author={Danilov, Don and Robledo-Moreno, Javier and Sung, Kevin J and Motta, Mario and Shee, James},
  journal = {J. Chem. Theory Comput.},
  year={2025},
  publisher={ACS Publications}
}

@article{barison2025quantum,
  title={Quantum-centric computation of molecular excited states with extended sample-based quantum diagonalization},
  author={Barison, Stefano and Moreno, Javier Robledo and Motta, Mario},
  journal = {Quantum},
  volume={10},
  number={2},
  pages={025034},
  year={2025},
  publisher={IOP Publishing}
}

@article{robledo2024chemistry,
  title={Chemistry beyond exact solutions on a quantum-centric supercomputer},
  author={Robledo-Moreno, Javier and Motta, Mario and Haas, Holger and Javadi-Abhari, Ali and Jurcevic, Petar and Kirby, William and Martiel, Simon and Sharma, Kunal and Sharma, Sandeep and Shirakawa, Tomonori and others},
  journal = {arXiv},
  year={2024}
}

@article{google2025observation,
  title={Observation of constructive interference at the edge of quantum ergodicity},
  journal={Nature},
  volume={646},
  number={8086},
  pages={825--830},
  year={2025},
  publisher={Nature Publishing Group UK London}
}

@article{babbush2025grand,
  title={The grand challenge of quantum applications},
  author={Babbush, Ryan and King, Robbie and Boixo, Sergio and Huggins, William and Khattar, Tanuj and Low, Guang Hao and McClean, Jarrod R and O'Brien, Thomas and Rubin, Nicholas C},
  journal = {arXiv},
  year={2025}
}

@article{bauer2020quantum,
  title={Quantum algorithms for quantum chemistry and quantum materials science},
  author={Bauer, Bela and Bravyi, Sergey and Motta, Mario and Chan, Garnet Kin-Lic},
  journal = {Chem. Rev.},
  volume={120},
  number={22},
  pages={12685--12717},
  year={2020},
  publisher={ACS Publications}
}

@article{lee2023evaluating,
  title={Evaluating the evidence for exponential quantum advantage in ground-state quantum chemistry},
  author={Lee, Seunghoon and Lee, Joonho and Zhai, Huanchen and Tong, Yu and Dalzell, Alexander M and Kumar, Ashutosh and Helms, Phillip and Gray, Johnnie and Cui, Zhi-Hao and Liu, Wenyuan and others},
  journal = {Nat. Commun.},
  volume={14},
  number={1},
  pages={1952},
  year={2023},
  publisher={Nature Publishing Group UK London}
}

@article{ramoa2025reducing,
  title={Reducing the resources required by ADAPT-VQE using coupled exchange operators and improved subroutines},
  author={Ram{\^o}a, Mafalda and Anastasiou, Panagiotis G and Santos, Luis Paulo and Mayhall, Nicholas J and Barnes, Edwin and Economou, Sophia E},
  journal = {Quantum},
  volume={11},
  number={1},
  pages={86},
  year={2025},
  publisher={Nature Publishing Group UK London}
}

@article{lin2025dissipative,
  title={Dissipative preparation of many-body quantum states: Toward practical quantum advantage},
  author={Lin, Lin},
  journal={APL Computational Physics},
  volume={1},
  number={1},
  pages={010901},
  year={2025},
  publisher={AIP Publishing LLC}
}

@article{cortes2022quantum,
  title={Quantum Krylov subspace algorithms for ground-and excited-state energy estimation},
  author={Cortes, Cristian L and Gray, Stephen K},
  journal = {Phys. Rev. A},
  volume={105},
  number={2},
  pages={022417},
  year={2022},
  publisher={APS}
}

@article{nandy2025quantum,
  title={Quantum dynamics in Krylov space: Methods and applications},
  author={Nandy, Pratik and Matsoukas-Roubeas, Apollonas S and Mart{\'\i}nez-Azcona, Pablo and Dymarsky, Anatoly and del Campo, Adolfo},
  journal={Physics Reports},
  volume={1125},
  pages={1--82},
  year={2025},
  publisher={Elsevier}
}

@article{kirby2023exact,
  title={Exact and efficient Lanczos method on a quantum computer},
  author={Kirby, William and Motta, Mario and Mezzacapo, Antonio},
  journal = {Quantum},
  volume={7},
  pages={1018},
  year={2023},
  publisher={Verein zur F{\"o}rderung des Open Access Publizierens in den Quantenwissenschaften}
}

@article{motta2024subspace,
  title={Subspace methods for electronic structure simulations on quantum computers},
  author={Motta, Mario and Kirby, William and Liepuoniute, Ieva and Sung, Kevin J and Cohn, Jeffrey and Mezzacapo, Antonio and Klymko, Katherine and Nguyen, Nam and Yoshioka, Nobuyuki and Rice, Julia E},
  journal={Electronic Structure},
  volume={6},
  number={1},
  pages={013001},
  year={2024},
  publisher={IOP Publishing}
}

@article{alexeev2025perspective,
  title={A perspective on quantum computing applications in quantum chemistry using 25--100 logical qubits},
  author={Alexeev, Yuri and Batista, Victor S and Bauman, Nicholas and Bertels, Luke and Claudino, Daniel and Dutta, Rishab and Gagliardi, Laura and Godwin, Scott and Govind, Niranjan and Head-Gordon, Martin and others},
  journal = {J. Chem. Theory Comput.},
  year={2025},
  publisher={ACS Publications}
}

@article{tkachenko2024quantum,
  title={Quantum Davidson algorithm for excited states},
  author={Tkachenko, Nikolay V and Cincio, Lukasz and Boldyrev, Alexander I and Tretiak, Sergei and Dub, Pavel A and Zhang, Yu},
  journal = {Quantum},
  volume={9},
  number={3},
  pages={035012},
  year={2024},
  publisher={IOP Publishing}
}

@article{lee2024sampling,
  title={Sampling error analysis in quantum krylov subspace diagonalization},
  author={Lee, Gwonhak and Lee, Dongkeun and Huh, Joonsuk},
  journal = {Quantum},
  volume={8},
  pages={1477},
  year={2024},
  publisher={Verein zur F{\"o}rderung des Open Access Publizierens in den Quantenwissenschaften}
}

@article{rohwedder2011analysis,
  title={An analysis for the DIIS acceleration method used in quantum chemistry calculations},
  author={Rohwedder, Thorsten and Schneider, Reinhold},
  journal={Journal of mathematical chemistry},
  volume={49},
  number={9},
  pages={1889--1914},
  year={2011},
  publisher={Springer}
}

@article{shen2023real,
  title={Real-time Krylov theory for quantum computing algorithms},
  author={Shen, Yizhi and Klymko, Katherine and Sud, James and Williams-Young, David B and de Jong, Wibe A and Tubman, Norm M},
  journal = {Quantum},
  volume={7},
  pages={1066},
  year={2023},
  publisher={Verein zur F{\"o}rderung des Open Access Publizierens in den Quantenwissenschaften}
}

@article{mcardle2019variational,
  title={Variational ansatz-based quantum simulation of imaginary time evolution},
  author={McArdle, Sam and Jones, Tyson and Endo, Suguru and Li, Ying and Benjamin, Simon C and Yuan, Xiao},
  journal = {Quantum},
  volume={5},
  number={1},
  pages={75},
  year={2019},
  publisher={Nature Publishing Group UK London}
}

@article{oumarou2025molecular,
  title={Molecular properties from quantum krylov subspace diagonalization},
  author={Oumarou, Oumarou and Ollitrault, Pauline J and Cortes, Cristian L and Scheurer, Maximilian and Parrish, Robert M and Gogolin, Christian},
  journal = {J. Chem. Theory Comput.},
  volume={21},
  number={9},
  pages={4543--4552},
  year={2025},
  publisher={ACS Publications}
}

@article{kitaev1995quantum,
  title={Quantum measurements and the Abelian stabilizer problem},
  author={Kitaev, A Yu},
  journal={arXiv preprint quant-ph/9511026},
  year={1995}
}

@article{parrish2019quantum,
  title={Quantum filter diagonalization: Quantum eigendecomposition without full quantum phase estimation},
  author={Parrish, Robert M and McMahon, Peter L},
  journal = {arXiv},
  year={2019}
}

@article{cohn2021quantum,
  title={Quantum filter diagonalization with compressed double-factorized hamiltonians},
  author={Cohn, Jeffrey and Motta, Mario and Parrish, Robert M},
  journal = {Quantum},
  volume={2},
  number={4},
  pages={040352},
  year={2021},
  publisher={APS}
}

@article{yoshioka2025krylov,
  title={Krylov diagonalization of large many-body Hamiltonians on a quantum processor},
  author={Yoshioka, Nobuyuki and Amico, Mirko and Kirby, William and Jurcevic, Petar and Dutt, Arkopal and Fuller, Bryce and Garion, Shelly and Haas, Holger and Hamamura, Ikko and Ivrii, Alexander and others},
  journal = {Nat. Commun.},
  volume={16},
  number={1},
  pages={5014},
  year={2025},
  publisher={Nature Publishing Group UK London}
}

@article{zhang2024measurement,
  title={Measurement-efficient quantum krylov subspace diagonalisation},
  author={Zhang, Zongkang and Wang, Anbang and Xu, Xiaosi and Li, Ying},
  journal={Quantum},
  volume={8},
  pages={1438},
  year={2024},
  publisher={Verein zur F{\"o}rderung des Open Access Publizierens in den Quantenwissenschaften}
}

@article{klymko2022real,
  title={Real-time evolution for ultracompact Hamiltonian eigenstates on quantum hardware},
  author={Klymko, Katherine and Mejuto-Zaera, Carlos and Cotton, Stephen J and Wudarski, Filip and Urbanek, Miroslav and Hait, Diptarka and Head-Gordon, Martin and Whaley, K Birgitta and Moussa, Jonathan and Wiebe, Nathan and others},
  journal={PRX Quantum},
  volume={3},
  number={2},
  pages={020323},
  year={2022},
  publisher={APS}
}

@article{seki2021quantum,
  title={Quantum power method by a superposition of time-evolved states},
  author={Seki, Kazuhiro and Yunoki, Seiji},
  journal={PRX Quantum},
  volume={2},
  number={1},
  pages={010333},
  year={2021},
  publisher={APS}
}

@article{yeter2020practical,
  title={Practical quantum computation of chemical and nuclear energy levels using quantum imaginary time evolution and Lanczos algorithms},
  author={Yeter-Aydeniz, K{\"u}bra and Pooser, Raphael C and Siopsis, George},
  journal={npj Quantum Information},
  volume={6},
  number={1},
  pages={63},
  year={2020},
  publisher={Nature Publishing Group UK London}
}

@misc{Sundar2026ChemicallyDecisive,
  title         = {Chemically decisive benchmarks on the path to quantum utility},
  author        = {Sundar, Srivathsan Poyyapakkam and Abraham, Vibin and Peng, Bo and Asthana, Ayush},
  year          = {2026},
  month         = jan,
  eprint        = {2601.10813},
  archivePrefix = {arXiv},
  primaryClass  = {physics.chem-ph},
  doi           = {10.48550/arXiv.2601.10813},
  url           = {https://arxiv.org/abs/2601.10813}
}

@article{schlimgen2022quantum,
  title={Quantum simulation of the Lindblad equation using a unitary decomposition of operators},
  author={Schlimgen, Anthony W and Head-Marsden, Kade and Sager, LeeAnn M and Narang, Prineha and Mazziotti, David A},
  journal={Physical Review Research},
  volume={4},
  number={2},
  pages={023216},
  year={2022},
  publisher={APS}
}
\end{document}